\documentclass[multphys,vecphys]{svmult}

\usepackage{makeidx}     
\usepackage{graphicx}    
\usepackage{multicol}    

\makeindex             

\begin{document}
\title*{GRB optical and IR rapid follow-up with the \\ 2 m Liverpool Robotic Telescope}
\titlerunning{GRB rapid follow-up with the Liverpool Telescope}
\author{Andreja Gomboc\inst{1,2}, Michael F. Bode\inst{1}, David Carter\inst{1}, 
Carole G. Mundell\inst{1}, Andrew M. Newsam\inst{1}, Robert J. Smith\inst{1}
\and Iain A. Steele\inst{1}}
\authorrunning{Gomboc et al.}
\institute{Astrophysics Research Institute, Liverpool John Moores University,
12 Quays House, Egerton Wharf, Birkenhead, CH41 1LD, United Kingdom,
\texttt{ag@astro.livjm.ac.uk, mfb@astro.livjm.ac.uk}
\and Department of Physics, University in Ljubljana, Jadranska 19, 1000 Ljubljana, Slovenia}

\maketitle

The Liverpool Telescope, owned and operated by Liverpool John Moores University and 
situated at Roque de los Muchachos, La Palma, is the first 2-m, fully instrumented robotic 
telescope. We plan to use the LT in conjunction with Gamma Ray Observatories (HETE-2, INTEGRAL, Swift) 
to study GRB physics. A special over-ride mode will enable observations commencing less than a minute 
after the GRB alert, including optical and near infrared imaging and spectroscopy. These observations, 
together with systematic monitoring of the burst through the afterglow, will help to unravel the 
nature of prompt optical flashes, short bursts, optically dark bursts, redshift distribution, 
GRB - supernova connection and other questions related to the GRB phenomenon.   In particular, 
the combination of aperture, instrumentation and rapid automated response makes the Liverpool 
Telescope excellently suited to the investigation of optically dark bursts and currently optically 
unstudied short bursts. 

\section{Introduction}

Early acquisition of multi-wavelength light curves for many GRBs is essential to further
our understanding of the nature and origin of these objects, the relationship between the 
prompt and afterglow emission and to distinguish between different afterglow models.
On the other hand, systematic observation of the afterglows for weeks following the GRB 
will help determine the connection between GRB and supernovae. 

Robotic operation of a telescope has the advantages of providing rapid reaction to short and 
unpredictable phenomena and their systematic follow-up, simultaneous or coordinated
with other ground facilities or satellites. This makes such facilities invaluable in 
the study of GRBs.

\begin{figure*}
\centering
\includegraphics[height=8cm]{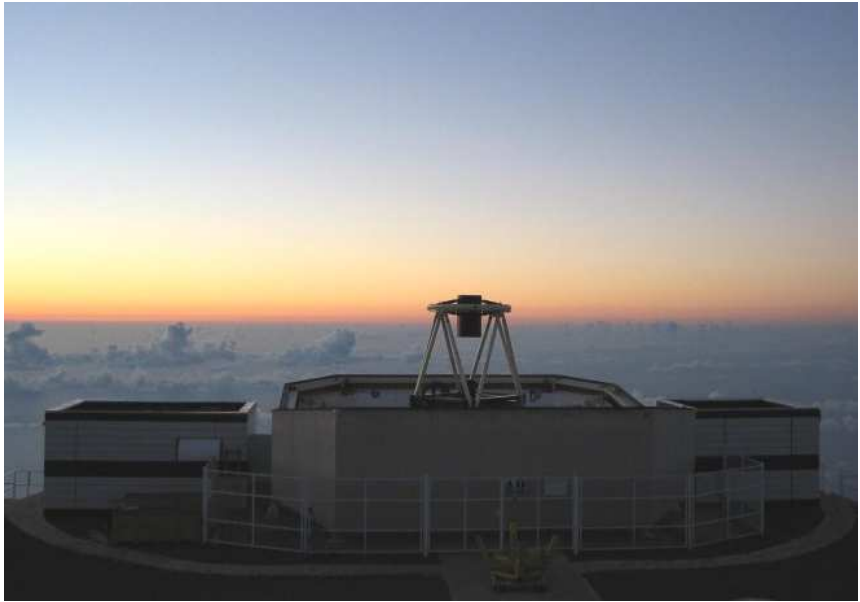}
\caption{The Liverpool Telescope at Roque de los Muchachos, La Palma, Canaries is 
a 2-m fully robotic altitude-azimuth design telescope. Fully openable enclosure 
and the slew rate $>$ 2$^{\rm o}$/s enable the start of observations in less than a minute
after the GRB alert.}
\label{figLT}       
\end{figure*}

The Liverpool Telescope (LT), with a primary mirror diameter of 2 m, is the largest
fully robotic telescope (Fig. \ref{figLT}). It is located on the excellent astronomical
site of La Palma in the Canaries and is housed in a unique fully-opening enclosure.
It can have 5 permanently mounted instruments, which are selected automatically by 
a deployable, rotating mirror in the A\&G box within 30s. 

\noindent
The LT instrumentation currently comprises 

{\it RATCam} Optical CCD Camera with 2048$\times$2048 pixels, FoV 4.6'$\times$4.6'
and 8 filter selections (u', g', r', i', z', B, V, ND2.0)

\noindent
and will be complemented with: 

{\it SupIRCam} 1-2.5 micron Camera with 256$\times$ 256 pixels, FoV 1.7'$\times$1.7'
and Z, J, H, K' filters - in Autumn 2003,

{\it Prototype Spectrograph} with 49, 1.7" fibres, 512$\times$512 pixels, R=1000;
  3500$<$$\lambda$$<$7000 \AA  - in Autumn 2003, and
  
 {\it FRODOSpec} Integral field double beam spectrograph with R=4000, 8000; 
 4000$<$$\lambda$$<$9500 \AA  - in Summer 2004.

\vspace{0.5cm}
 First light on the LT was successfully achieved at the end of July 2003 and currently, 
 the LT is still in the commissioning phase. For updated information please see  
 http://telescope.livjm.ac.uk/.

\section{GRB follow-up with the LT}
GRB follow-up observations have been assigned high priority in the LT observing
programme. Following the GRB alert from the GCN network and HETE-2, INTEGRAL and Swift,
we will employ the Over-Ride mode  and commence the search for and observation of GRB 
counterparts. The initial LT strategy for responding to Swift alerts is presented in 
Table \ref{tableLT}. With this routine as the starting point, the automated procedures 
can be optimised with experience and adapted regarding scientific imperatives.

\begin{table}
\centering
\caption{Initial LT strategy for responding to Swift alerts.
}
\label{tab:1}      
\begin{tabular}{lll}
\hline\noalign{\smallskip}
Time (s)  \hspace{0.5cm} & Swift instrument and the  \hspace{0.5cm} & Alert progress and   \\
 & error box of GRB position & simultaneous LT activity \\
\noalign{\smallskip}\hline\noalign{\smallskip}
0 &  & GRB alert \\
15 & BAT ($\gamma$-ray) $\sim$ 4' & Release of GCN alert \\
20 & & GCN alert arrives at LT - Automatic \\
& & override starts slew \\
50 & & Commence multiband optical imaging \\
& & (FOV=4.6') \\
140 & XRT (X-ray) $\sim$ 5" & Re-point and select SupIRCam \\
& & (FOV=1.7') \\
160 & & Commence near infrared imaging \\
320 & UVOT (opt.) $\sim$ 0.3" & Re-point and select spectrograph \\
& & (FOV=10") \\
340 & & Commence optical spectroscopy \\
\noalign{\smallskip}\hline
\end{tabular}
\label{tableLT}
\end{table}

\section{GRB science with the LT}

\subsection{Short bursts}
Of all the GRB afterglows so far observed, only one single epoch observation \cite{Castro-Tirado}
of a possible optical counterpart to a short GRB was reported. As it may be expected 
that short GRB afterglows will be 3-4 magnitudes fainter than long GRB afterglows \cite{Panaitescu},
a 2-m or even larger telescope is required for their rapid follow-up, or, in the case of no detection,
to place even more stringent limits on the afterglow immediately following the short burst.

\subsection{Spectroscopy and redshift}
Redshift information is currently available for about 40 long bursts \cite{jcg} with the selection 
likely to be extremely biased towards the brightest events, those with the slowest optical decay 
curves or those with the brightest host galaxies. In the search for high-redshift GRBs, infra-red 
imaging is particularly important. Lyman limit absorption will heavily extinguish the optical 
emission from objects with z$>$10. In the near infrared however, Ly$\alpha$ emission would still 
be clearly detected. A near-infrared detection of an optically dark GRB is the signature of either 
a very high redshift or highly dust enshrouded event.

Spectroscopy also has the potential to probe the evolution of the circumstellar and interstellar
environment of the burst. Currently, due to lack of spectra obtained in the crucial early
phases of the bursts, this is largely unexplored territory.

\subsection{Prompt flashes}
Currently there are only 3 GRBs (GRB990123, GRB021004 and GRB021211) 
with optical afterglows detected within the first ten minutes after the GRB initial event, 
of which GRB021004 has only a poorly sampled light curve. The prompt flashes following GRB990123 
and GRB021211 have been extensively analyzed \cite{Akerlof}, \cite{Li} and show similar rapid 
decay rates (3-5 magnitudes in 10 min) despite GRB990123 being about four magnitudes brighter 
at peak (R$_{peak}\sim$9) than GRB021211 (R$_{peak}<$14). Given this rapid decline, it is easy 
to imagine that the roughly 50\% of the bursts currently considered optically 'dark', may be 
detected by a larger rapid reaction telescope such as the LT. Furthermore, in both the above 
cases, interpretation is hindered by there only being white-light unfiltered observations. On 
the basis of these though, evidence has been cited for rapid colour changes during the first 
minute \cite{Li}, underlining the need on these time scales for multicolour filtered photometry.
The limiting magnitude (V=22) and photometric accuracy of the LT will allow direct and detailed 
measurement of prompt optical and infrared flashes.

\subsection{Burst physics}
Existing afterglow observations are broadly consistent with fireball models,
but more complex models involving beaming, jets and disks are not well constrained by 
existing data \cite{Frail}, \cite{Piran}, \cite{Rossi}. High quality light curves and spectra, 
which track the source evolution from initial burst stages through to late afterglow probing 
both the energetics of the progenitor and its interaction with its surrounding will help to 
determine several important parameters. These include the identity of the progenitors, the 
nature of triggering mechanism, the physics of the energy transport during burst and afterglow, 
the timescales involved and the interaction between the ejected material and the surrounding medium.

\subsection{Supernova connection}
The first evidence of a possible association of GRBs with supernovae was reported for the
GRB 980425 and SN1998bw \cite{Galama}. The recent discovery of temporal and spatial coincidence 
of GRB 030329 and SN2003dh together with the spectral evolution \cite{Hjorth} gives significant 
support to the GRB-supernova connection and hence the model that long duration GRBs originate in 
the death (core collapse) of massive stars. Systematic observations of afterglows also at later 
stages, 10-30 days after the GRB, are therefore essential to reveal more about the GRB phenomena 
and their link to supernovae.

\section{Conclusions}
Rapid response time, moderate aperture, excellent site and range of instrumentation make the 
LT especially suitable for study of afterglows of short GRBs, afterglows of optically dark GRBs, 
prompt optical flashes, early afterglow spectrometry and statistical properties of GRBs and their 
afterglows. With approximately 25\% of GRBs occuring at night over La Palma and 70$^{\rm{o}}$ maximal 
zenith distance observable by the LT, we expect to observe 1 in 6 GRBs immediately following 
the alert. We plan to monitor GRB afterglows also at later stages depending on their scientific 
significance and in collaboration with other facilities, including the Faulkes Telescopes 
(clones of LT sited in Australia and Hawaii).

\vspace{0.5cm}
{\it Acknowledgments} \\
The Liverpool Telescope is funded via EU, PPARC, JMU grants and the benefaction of Mr. A. E. Robarts.
A.G. acknowledges the receipt of the Marie Curie Fellowship from the EU.

\printindex
\end{document}